\documentclass[12pt]{article}
\usepackage{epsf}
\usepackage{graphicx}

\begin{document}
\title
{Super-strong interacting gravitons as a main engine of the
universe without expansion or dark energy}
\author
{Michael A. Ivanov \\
Physics Dept.,\\
Belarus State University of Informatics and Radioelectronics, \\
6 P. Brovka Street,  BY 220027, Minsk, Republic of Belarus.\\
E-mail: ivanovma@gw.bsuir.unibel.by.}

\maketitle

\begin{abstract}
The basic cosmological conjecture about the Dopplerian nature of
redshifts may be false if gravitons are super-strong interacting
particles. A quantum mechanism of classical gravity and the main
features of a new cosmological paradigm based on it are described
here.
\end{abstract}
%PACS 04.60.-m, 98.70.Vc

%\section[1]{The graviton background and the Newtonian limit}

If we assume that the background of super-strong interacting
gravitons  exists, then the classical gravitational attraction
between any pair of bodies arises due to a Le Sage's kind
mechanism. A net force of attraction and repulsion will be
non-zero if one suggests that graviton pairs exist and these pairs
are destructed by collisions. This pairing is like to the one
having place in a case of superconductivity. The portion of
pairing gravitons, $2\bar n_{2}/\bar n,$  a spectrum of single
gravitons, $f(x),$ and a spectrum of subsystem of pairing
gravitons, $f_{2}(2x),$ are shown on Fig. 1 as functions of the
dimensionless parameter $x \equiv \hbar \omega /kT$ (for more
details, see \cite{1}).
\par
\begin{figure}[th]
\centerline{\includegraphics[width=12.98cm]{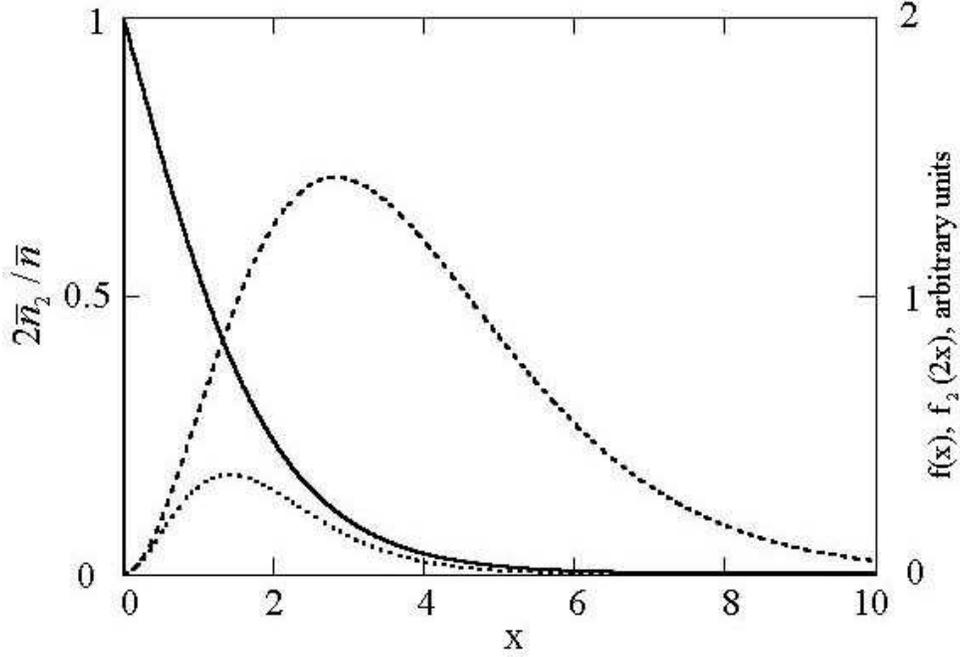}}
\caption{The portion of pairing gravitons, $2\bar n_{2}/\bar n,$
(solid line), a spectrum of single gravitons, $f(x),$ (dashed
line), and a spectrum of graviton pairs, $f_{2}(2x),$ (dotted
line) versus the dimensionless parameter $x$.}
\end{figure}
By the Planckian spectra of gravitons we find for the Newtonian
constant \cite{1}:
\begin{equation}
G = {2 \over 3} \cdot {D^{2} c(kT)^{6} \over {\pi^{3}\hbar^{3}}}
\cdot I_{2}
\end{equation}
where $I_{2}=2.3184 \cdot 10^{-6},$ $T$ is an effective
temperature of the background, and $D$ is some new dimensional
constant. It is necessary to accept for a value of this constant:
$D=1.124 \cdot 10^{-27}{m^{2} / eV^{2}}.$

%\section[2]{Redshifts and supernovae dimming}

In a presence of the graviton background, which is considered in a
flat space-time, an energy of any photon should decrease with a
distance $r,$ so we have for a redshift $z$ \cite{2}:
$z=\exp(ar)-1,$ where $a=H/c$ with the Hubble constant:
\begin{equation}
H= {1 \over 2\pi} D \cdot \bar \epsilon \cdot (\sigma T^{4}),
\end{equation}
where $\bar \epsilon$ is an average graviton energy, $\sigma$ is
the Stephan-Boltzmann constant.
\par
It means that in this approach the two fundamental constants, $G$
and $H,$ are connected between themselves:
\begin{equation}
H= (G  {45 \over 32 \pi^{5}}  {\sigma T^{4} I_{4}^{2} \over
{c^{3}I_{2}}})^{1/2},
\end{equation}
with $I_{4}=24.866.$ Using the known value of $G,$ one can
compute:$H= 3.026 \cdot 10^{-18}s^{-1}=94.576 \ km \cdot s^{-1}
\cdot Mpc^{-1}$ by $T=2.7 K.$
\par
From another side, an additional relaxation of any photonic flux
due to non-forehead collisions of gravitons with photons leads to
a luminosity distance $D_{L}:$
\begin{equation}
D_{L}=a^{-1} \ln(1+z)\cdot (1+z)^{(1+b)/2} \equiv a^{-1}f_{1}(z),
\end{equation}
where $b= 3/2+2/\pi =2.137.$
\par
\begin{figure}[th]
\centerline{\includegraphics[width=12.98cm]{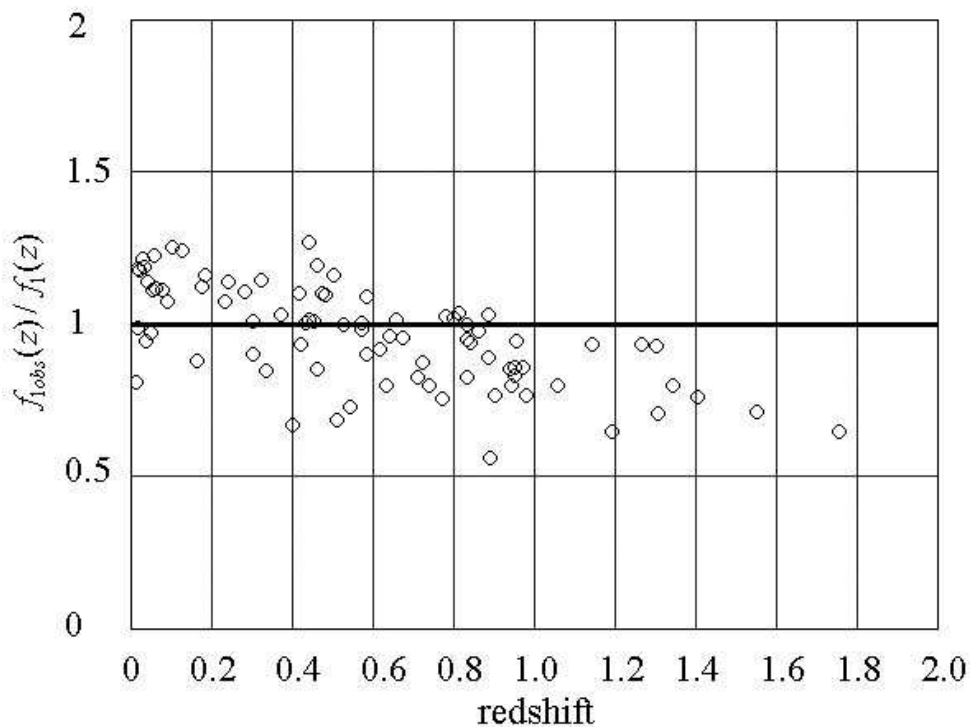}}
\caption{The ratio of observed to theoretical functions $f_{1
obs}(z)/f_{1}(z)$ (dots); observational data are taken from Table
5 of \cite{3}. If this model is true, the ratio should be equal to
1 for any $z$ (solid line).}
\end{figure}
This function fits supernovae data well for $z < 0.5$ \cite{4}. It
excludes a need of any dark energy to explain supernovae dimming.
If one introduces distance moduli $\mu_{0} = 5 \log D_{L} + 25 = 5
\log f_{1 obs} + c_{1}$, where $c_{1}$ is a constant, $f_{1
obs}(z)$ is an observed analog of $f_{1}(z)$, we can compute the
ratio $f_{1 obs}(z)/f_{1}(z)$ using recent supernovae
observational data from \cite{3} (see Fig. 2).
\par
The question arises: how are gravitons and photons connected? If
the conjecture by Adler et al. \cite{a98} (that a graviton with
spin $2$ is composed with two photons) is true, one might check it
in a laser experiment. Taking two lasers with photon energies
$h\nu_{1}$ and $h\nu_{2}$, one may force laser beams to collide on
a way $L$ (see Fig. 3). If photons self-interact, then photons
with energies $h\nu_{1}-h\nu_{2}$, if $h\nu_{1}>h\nu_{2}$, would
arise after collisions of initial photons.
\begin{figure}[th]
\centerline{\includegraphics[width=12.98cm]{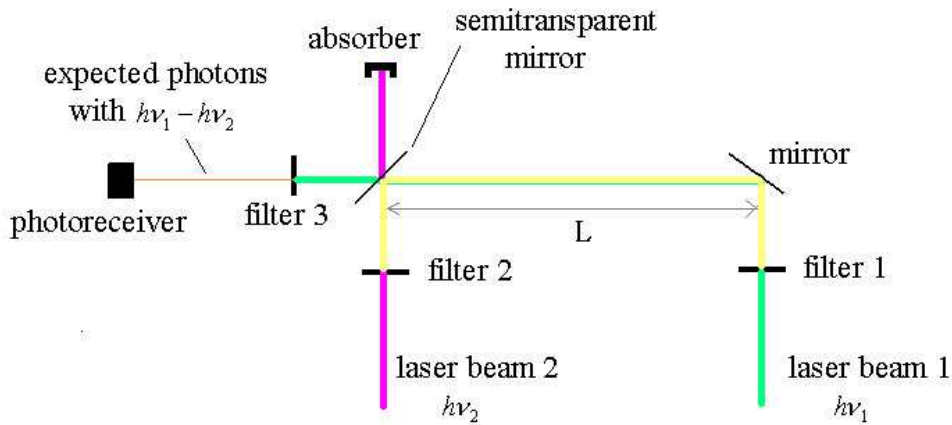}}
\caption{The scheme of laser beam passes.}
\end{figure}
If we {\it assume} that single gravitons are identical to photons,
then an average graviton energy should be replaced with
$h\nu_{2}$, the factor $1/2\pi$ in (2) should be replaced with
$1/\varphi$, where $\varphi$ is a divergence of laser beam 2, and
one must use a quantity $P/S$ instead of $\sigma T^{4}$ in (2),
where $P$ is a laser 2 power and $S$ is a cross-section of its
beam. It means that we should replace the Hubble constant with its
analog for a laser beam collision, $H_{laser}$: $H \rightarrow
H_{laser} = {1 \over \varphi} \cdot D \cdot h\nu_{2}\cdot {P \over
S}.$ Taken $\varphi=10^{-4}$, $h\nu_{2} \sim 1~eV$, $P \sim
10~mW$, and $P/S \sim 10^{3}~W/m^{2}$, that is characterizing a
He-Ne laser, we get: $H_{laser} \sim 0.06 ~s^{-1}$. Then photons
with energies $h\nu_{1}-h\nu_{2}$ would fall to a photoreceiver
with a frequency which should linearly rise with $L$, and it would
be of $10^{7}~s^{-1}$ if both lasers have equal powers $\sim
10~mW$, and $L\sim 1~m$. It is a big enough frequency to detect
easy a flux of these expected photons in the IR band.
\par
In this approach (its summarizing description \cite{77} will be
soon published), every massive body would be decelerated due to
collisions with gravitons \cite{2} that may be connected with the
Pioneer 10 anomaly \cite{33}.

\end{document}